\documentclass[letterpaper, 10 pt, journal,twoside]{IEEEtran}  

\IEEEoverridecommandlockouts                              

\usepackage{graphics} 
\usepackage{amsmath} 
\usepackage{amssymb}  
\usepackage{graphicx}
\usepackage{algorithm}
\usepackage{cite}
\usepackage[noend]{algpseudocode}
\usepackage{hyperref}
\usepackage{subfloat}
\usepackage{tikz}
\usepackage{textcomp}
\usepackage{lipsum}

\DeclareMathOperator*{\argmin}{\arg\!\min}

\algnewcommand{\IfThenElse}[3]{
  \State \algorithmicif\ #1\ \algorithmicthen\ #2\ \algorithmicelse\ #3}
  
\newcommand\copyrighttext{%
  \scriptsize \textcopyright 2019 IEEE. Personal use of this material is permitted. Permission from IEEE must be obtained for all other uses, in any current or future media, including reprinting/republishing this material for advertising or promotional purposes, creating new collective works, for resale or redistribution to servers or lists, or reuse of any copyrighted component of this work in other works.
}
\newcommand\copyrightnotice{%
\begin{tikzpicture}[remember picture,overlay]
\node[anchor=south,yshift=10pt] at (current page.south) {\fbox{\parbox{\dimexpr\textwidth-\fboxsep-\fboxrule\relax}{\copyrighttext}}};
\end{tikzpicture}%
}

\title{Hierarchical Event-triggered Learning for Cyclically Excited Systems with Application to Wireless Sensor Networks*
}

\author{Jonas Beuchert$^{1}$, Friedrich Solowjow$^{2}$, J\"org Raisch$^{1}$, Sebastian Trimpe$^{2}$, and Thomas Seel$^{1}$
\thanks{*The work of F. Solowjow and S. Trimpe was supported in part by the Max Planck Society, the IMPRS-IS, and the Cyber Valley Initiative.}
\thanks{$^{1}$Jonas Beuchert, J\"org Raisch, and Thomas Seel are with Control Systems Group, Technische Universit\"{a}t Berlin, 10587 Berlin, Germany
        {\tt\small jonas.beuchert@campus.tu-berlin.de, seel@control.tu-berlin.de}.}%
\thanks{$^{2}$Friedrich Solowjow and Sebastian Trimpe are with Intelligent Control Systems Group, Max Planck Institute for Intelligent Systems, 70569 Stuttgart, Germany
        {\tt\small solowjow@is.mpg.de, trimpe@is.mpg.de}.}
}

\begin{document}

\maketitle
\thispagestyle{empty}
\pagestyle{empty}
\copyrightnotice

\begin{abstract}
Communication load is a limiting factor in many real-time systems. 
Event-triggered state estimation and event-triggered learning methods reduce network communication by sending information only when it cannot be adequately predicted based on previously transmitted data. 
This paper proposes an event-triggered learning approach for nonlinear discrete-time systems with cyclic excitation. The method automatically recognizes cyclic patterns in data -- even when they change repeatedly -- and reduces communication load whenever the current data can be accurately predicted from previous cycles. 
Nonetheless, a bounded error between original and received signal is guaranteed. The cyclic excitation model, which is used for predictions, is updated hierarchically, i.e., a full model update is only performed if updating a small number of model parameters is not sufficient. A nonparametric statistical test enforces that model updates happen only if the cyclic excitation changed with high probability. The effectiveness of the proposed methods is demonstrated using the application example of wireless real-time pitch angle measurements of a human foot in a feedback-controlled neuroprosthesis. The experimental results show that communication load can be reduced by 70\;\% while the root-mean-square error between measured and received angle is less than 1$^{\circ}$.

\end{abstract}


\section{Introduction}

\IEEEPARstart{M}{any} applications require real-time transmission of signals over communication channels with bandwidth limitations. A typical example is given by wireless sensor networks in feedback-controlled systems. The number of agents (i.e., network nodes) and their communication rate is limited by the amount of information the wireless network can transmit in real-time. It is, therefore, desirable to reduce the communication load without compromising the accuracy of the transmitted signals.

\begin{figure}[t]
      \centering
      \includegraphics[width=\columnwidth]{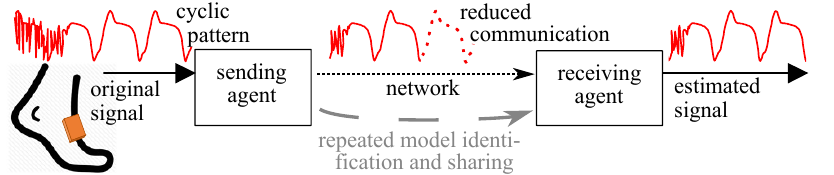}
      \caption{Event-triggered learning in a two-agents network. If the measured signal (e.g., a human foot angle) can be described by a cyclically excited system model, the model is shared with the receiver and the communication is reduced to those samples that cannot be predicted using the model and previously transmitted data. Inaccuracy of the model (e.g., due to change in walking pattern) is detected and a new model is identified and shared.}
      \label{fig:ETSE}
   \end{figure}

Well known approaches are event-based sampling \cite{c1,t1,t2} and event-triggered state estimation (ETSE \cite{t3,t4,t5}, sometimes referred to as model-based event-based sampling \cite{c2}): 
At each sampling instant, the receiving agent independently predicts the state, which is measured by the sender, based on previous estimations and a model. The sender performs the identical prediction and communicates the measured state if and only if the error between prediction and measurement exceeds a predefined threshold. Otherwise, there is no communication and the receiving agent uses the model-based prediction as estimation (cf. Fig.~\ref{fig:ETSE}).

Since the prediction accuracy heavily depends on the quality of the utilized model, it was recently proposed to learn and update models in an event-triggered fashion as well \cite{c4}. Occurrence of communication is treated as a random variable, and the model is updated when empirical data does not fit the probability distribution that would result if the model was the truth.

The present paper builds on the idea of \cite{c4} and develops event-triggered learning (ETL) methods for cyclically excited systems. The main contributions are:
\begin{itemize}
 \item Extension of the concept of ETL to specifically target systems with a locally cyclic excitation. Locally means that cycles close in time are almost identical; however, cycles that are far apart are not necessarily similar. This class of systems is useful for describing both biological and technical processes (e.g., human motion, breath, heartbeat, and production cycles).
 \item Utilization of a learning trigger tailored to the problem at hand (one-sided Kolmogorov-Smirnov test \cite{ksTest}), which fires with high probability in case of a model change and with low probability otherwise.
 \item Introduction of the novel idea of a hierarchical model learning strategy, which updates and communicates only a reduced number of model parameters whenever that is sufficient.
 \item Demonstration of significant communication savings (70~\%) using experimental data from cyclic human motion collected with a wearable inertial sensor network. This is the first application of ETL on real-world network data.

\end{itemize}


  \begin{figure*}[thpb]
      \centering
      \includegraphics[width=0.9\textwidth]{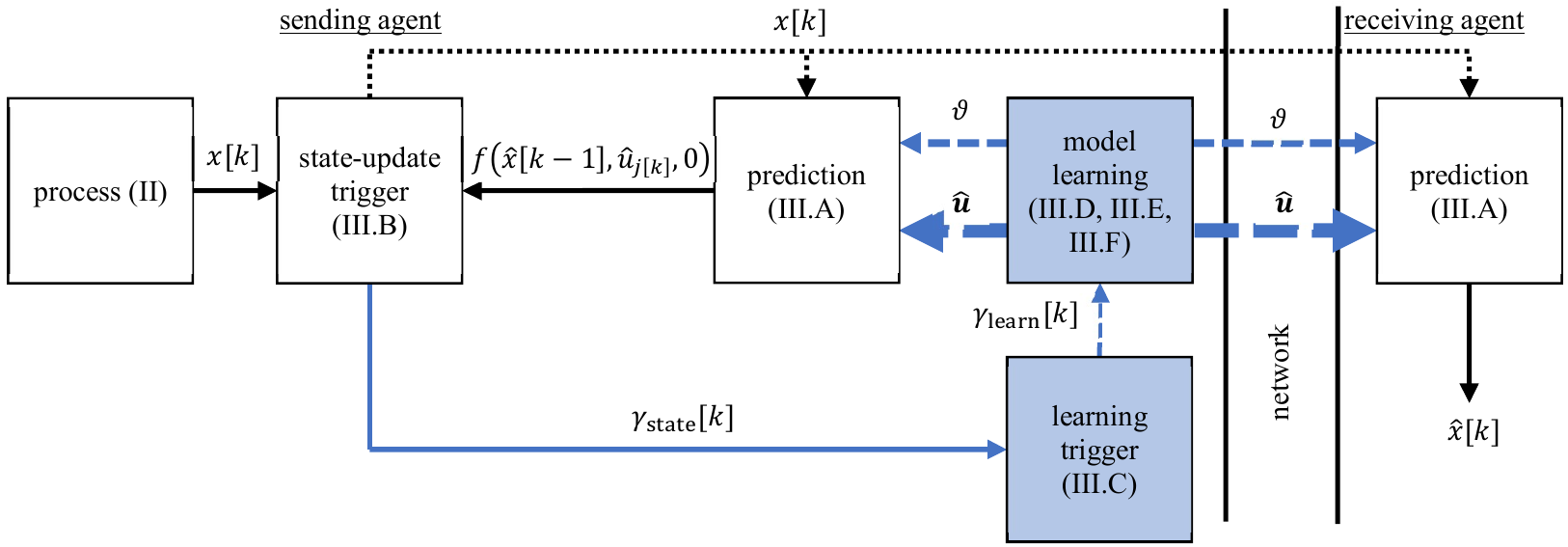}
      \caption{One sending and one receiving agent with the typical event-triggered state estimation architecture in black and event-triggered learning in blue (extended from \cite{c4}). The process provides the measured state $x[k]$ at every sampling instant $k$. At the same time, the state is estimated by the prediction blocks of the sender and the receiver using the previous estimate $\hat{x}\left[k-1\right]$ and a trajectory model of the excitation $\hat{\boldsymbol{u}}$. If the prediction differs significantly from the measured state, then a state update is triggered, and the internal states of both prediction blocks are set to the measured state. Too frequent state updates indicate poor model quality and, therefore, trigger model learning. This can either lead to only an adjustment of certain parameters $\vartheta$ of the current model trajectory or to a completely new excitation trajectory $\hat{\boldsymbol{u}}$. In any case, the new model information is shared between sender and receiver.}
      \label{fig:ETL}
   \end{figure*}

The paper continues as follows.
After defining the considered problem in Sec.~\ref{sec:problem}, the event-triggered learning architecture from \cite{c4} is briefly explained in Sec.~\ref{sec:ETL} and then extended to cyclically excited systems. Subsequently, the properties of the method are validated experimentally in Sec.~\ref{sec:eval}. Finally,  Sec.~\ref{sec:conclusions} provides conclusions.


\section{System Representation and Problem Formulation} \label{sec:problem}

Consider a discrete-time system with sample index $k\in \mathbb{Z}$ and state $x\left[k\right]\in \mathbb{R}^{n}$, which is measured by the sending agent. The system is assumed to be influenced by a cyclic input $u\left[k\right]\in \mathbb{R}^{m}$ with cycle length $N\in \mathbb{N}^{+}$ such that $u\left[k+N\right]=u[k]$ and by zero-mean noise $\varepsilon \left[k\right]\in \mathbb{R}^{s}$, which is distributed according to a time-invariant probability distribution $\mathcal{P}$. The recursive state update law of the system is characterized by the dynamics $f\colon \mathbb{R}^{n}\times \mathbb{R}^{m}\times \mathbb{R}^{s}\rightarrow \mathbb{R}^{n}$, i.e.,
\begin{equation}
\label{eq:sys}
x\left[k\right]=f\left(x\left[k-1\right],u\left[k\right],\varepsilon \left[k\right]\right).
\end{equation}
While the dynamics $f$ is assumed to be known, the excitation $u\left[k\right]$ and its cyclicity $N$ are unknown and may change with time. In the following, whenever the term \emph{model} is used, it refers to an approximation of the cyclic excitation $u\left[k\right]$. The distribution $\mathcal{P}$ of the noise $\varepsilon \left[k\right]$ is known.

This paper considers architectures with one sending and one receiving agent. However, the methods can be directly applied to multi-agent systems and yield the same advantages therein. 

The sending and receiving agents have the following capabilities, which will be made precise in the next section:
\begin{itemize}
 \item the sender can transmit measured data samples to the receiver, i.e., perform \emph{state updates};
 \item sender and receiver can estimate current data samples from previously transmitted data and a model, i.e., perform \emph{predictions};
 \item the sender can estimate excitation trajectories from measured data, i.e., perform \emph{model identification};
 \item the sender can send model parameters to the receiver, i.e., perform \emph{model updates}.
\end{itemize}
The main objective is to find a joint strategy for the sender and the receiver such that the amount of communication (state and model updates) is reduced while the error between the actual measurement signal $x\left[k\right]$ and the signal estimate $\hat{x}\left[k\right]\in \mathbb{R}^{n}$ on the receiver side remains small in the sense of a suitable metric $\left\| \cdot ,\cdot \right\| \colon \mathbb{R}^{n}\times \mathbb{R}^{n}\rightarrow \mathbb{R}_{0}^{+}$. 

\section{Algorithm Design} \label{sec:ETL}

The proposed event-triggered learning approach for cyclically excited systems is illustrated in Fig.~\ref{fig:ETL}. It can be described by the following building blocks:
\begin{itemize}
\item Two identical predictors that estimate the measured state $x[k]$ based on an internal state $\hat{x}[k-1]$ and an estimated excitation trajectory $\hat{\boldsymbol{u}}\in \mathbb{R}^{m\times \hat{N}}$ of one cycle with estimated cycle length $\hat{N}\in \mathbb{N}^{+}$.
\item A binary state-update trigger $\gamma _{\text{state}}\left[k\right]\in \left\{0,1\right\}$ that determines when to update the internal state $\hat{x}[k]$ of the predictors with the measurement $x\left[k\right]$ to ensure a bounded error $\left\| x\left[k\right],\hat{x}\left[k\right]\right\| $ of the estimation.
\item A model learning block that estimates the excitation trajectory $\hat{\boldsymbol{u}}$ used by the predictors or updates the estimated parameters $\vartheta\in \mathbb{R}^{v }$ (e.g., cycle length or amplitude) of the current trajectory $\hat{\boldsymbol{u}}$.
\item A binary learning trigger $\gamma _{\text{learn}}\left[k\right]\in \left\{0,1\right\}$ that determines when to update the internal excitation trajectory model $\hat{\boldsymbol{u}}$ of the predictors. Ideally, learning shall be triggered if and only if the rate of state updates increases due to a false or inaccurate model $\hat{\boldsymbol{u}}$. 
\end{itemize}

\subsection{Prediction} \label{sec:pred}


Let the current model of the cyclic excitation $u\left[k\right]$  be described by the aforementioned matrix $\hat{\boldsymbol{u}}$ such that the current excitation is the $(j[k])^\text{th}$ column of $\hat{\boldsymbol{u}}$, i.e., $\hat{u}_{j\left[k\right]}$ is the estimated value of $u\left[k\right]$, where the index $j\left[k\right]$ obeys
\begin{equation}
\label{eq:index}
j\left[k+1\right]=\begin{cases}
 j\left[k\right]+1 & \text{if }\gamma _{\text{learn}}\left[k\right]=0\wedge j\left[k\right]<\hat{N}\\
1 & \text{if }\gamma _{\text{learn}}\left[k\right]=1\vee j\left[k\right]=\hat{N}
\end{cases}
 .
\end{equation}
The index $j\left[k\right]$ is increased after every prediction step, unless the end of the estimated excitation trajectory is reached ($j\left[k\right]=\hat{N}$) or learning is triggered ($\gamma _{\text{learn}}\left[k\right]=1$).



The estimate $\hat{x}\left[k\right]$ is determined by
\begin{equation}
\label{eq:pred}
\hat{x}\left[k\right]=\begin{cases}
 f\left(\hat{x}\left[k-1\right],\hat{u}_{j\left[k\right]},0\right) & \text{if }\gamma _{\text{state}}\left[k\right]=0\\
x\left[k\right] & \text{if }\gamma _{\text{state}}\left[k\right]=1
\end{cases}
 .
\end{equation}

\subsection{State-update Trigger}

If $d\left[k\right]=\left\| x\left[k\right],f\left(\hat{x}\left[k-1\right],\hat{u}_{j\left[k\right]},0\right)\right\| $ reaches or exceeds the predefined threshold $\delta $, then a state update as defined in  Sec.~\ref{sec:pred} is triggered
\begin{equation}
\label{eq:su-trigger}
\gamma _{\text{state}}\left[k\right]=\begin{cases}
 0 & \text{if }d\left[k\right]<\delta \\
1 & \text{if }d\left[k\right]\geq \delta 
\end{cases}
 .
\end{equation}
Triggers that compare the actual value to a model-based prediction are common in ETSE (cf. \cite{c3,c5,c1,c2,trigger,t3,t4,t5}). Because a state update ($\gamma_{\text{state}}[k]=1$) leads to zero error, \eqref{eq:pred} and \eqref{eq:su-trigger} together ensure that $\left\| x\left[k\right],\hat{x}\left[k\right]\right\|$ is bounded by $\delta$.

\subsection{Robust Learning Trigger} \label{sec:l-trigger}

If the model (the estimated excitation trajectory $\hat{\boldsymbol{u}}$) is exact, then state updates occur only due to noise $\varepsilon\left[k\right]$. If the model no longer yields valid predictions of the current data, the time durations between two consecutive state updates will decrease. The learning trigger aims at detecting this decrease and then triggering a model update. 

Let the inter-communication time be the discrete time duration (number of samples) between two consecutive state updates and collect these times $\tau _{1},\tau _{2},\ldots ,\tau _{i}$ in a buffer, which is emptied whenever model learning is triggered ($\gamma _{\text{learn}}\left[k\right]=1$).


Perform Monte Carlo (MC) simulations of \eqref{eq:sys}, \eqref{eq:pred}, and \eqref{eq:su-trigger} for the case in which the model $\hat{\boldsymbol{u}}$ is perfect and state updates occur only due to noise. This yields a hypothetical cumulative distribution function $F\colon \mathbb{N}^{+}\rightarrow \left[0,1\right]$ of the inter-communication times.
At each sample instant $k$, the aforementioned buffer provides 
an empirical cumulative distribution function $\hat{F}_{k}\colon \mathbb{N}^{+}\rightarrow \left[0,1\right]$ of the previously observed inter-communication times; 
i.e., for any value $\tau\in\mathbb{N}^{+}$, $\hat{F}_{k}(\tau )$ is the proportion of observed inter-communication times less than or equal to $\tau $. 

The hypothetical and the empirical distribution function of the inter-communication times are compared by the one-sided two-sample Kolmogorov-Smirnov test (KS-test) \cite{ksTest, ksTestTab, ksTestFormula}. Its null hypothesis is that the empirical inter-communication times come from the hypothetical distribution $F(\tau )$. The alternative hypothesis is that they come from a different one. The result of the test is an estimated probability $p\left[k\right] \in \left[0,1\right]$ 
that the null hypothesis is true. It is compared with a defined significance level $\eta \in \left[0,1\right]$, i.e., the probability that the test rejects the null hypothesis although it is correct (type I error). If the $p\left[k\right]$-value is smaller than $\eta$ for a predefined minimum holding time duration $t_{\min }\in \mathbb{R}_{0}^{+}$, then a model update is triggered
\begin{equation}
\label{eq:l-trigger}
\gamma _{\text{learn}}\left[k\right]=\begin{cases}
 0 & \text{if }\exists \; l\in \left[k-t_{\min },k\right]\colon p\left[l\right] \geq \eta \\
1 & \text{if } p\left[l\right] < \eta \; \forall \; l\in \left[k-t_{\min },k\right]
\end{cases} 
 .
\end{equation}

Note that the null hypothesis is always accepted if no inter-communication times were observed since the last model update. Furthermore, for a large class of systems, the hypothetical distribution $F\left(\tau \right)$ will not depend on $\hat{\boldsymbol{u}}$ and, therefore, can be determined beforehand, i.e., no MC simulations must be performed in real time. Finally, for sufficiently simple systems, the distribution might even be determined analytically, and the one-sided one-sample KS-test can be used.

Utilizing the KS-test to design learning triggers was first proposed in \cite{newPaper}. Density-based learning triggers use richer statistical information and have more advantageous properties than learning triggers that are based on the expected value as proposed in \cite{c4}.
In contrast to \cite{newPaper}, the current method uses a one-sided trigger condition because model updates are not required if the inter-communication times are larger than expected. The robustification with a minimum holding time $t_{\min }$ prevents model learning due to unmodeled short-term effects. Furthermore, if a change of the process behavior occurs, then learning should not be triggered before the change is completed. For the theoretical properties of the statistical tests to hold, $\tau _{1},\tau _{2},\ldots ,\tau _{i}$ are assumed independent and identically distributed (see \cite{c4,newPaper}). While this is not necessarily the case for every system of the form \eqref{eq:sys}, it holds, for example, for the application system considered in Sec. \ref{sec:eval}.


\subsection{Model Learning -- Small Model Update}
\label{sec:small}

The model learning block learns a new model trajectory $\hat{\boldsymbol{u}}$ based on the previous measurements $x\left[k\right],x\left[k-1\right],\ldots $ However, transferring the complete trajectory $\hat{\boldsymbol{u}}$ of a cycle from the sending to the receiving agent leads to a significant amount of communication. Therefore, it is assumed that, in some cases, an appropriate new model trajectory $\hat{\boldsymbol{u}}_{+}\in \mathbb{R}^{m\times \hat{N}_{+}}$ with a new cycle length $\hat{N}_{+}$ can be derived from the old one by employing a parametric deformation function $g\colon \mathbb{R}^{m\times \hat{N}}\times \mathbb{R}^{v }\rightarrow \mathbb{R}^{m\times \hat{N}_{+}}$ with a small number of parameters $\vartheta\in \mathbb{R}^{v }$ such that $\hat{\boldsymbol{u}}_{+}=g\left(\hat{\boldsymbol{u}},\vartheta\right)$. 

Whenever a learning update is triggered, the learning block determines the parameters $\vartheta$ that lead to the best approximation of the states measured during the last cycle:
\begin{equation}
\label{eq:cost}
e(k,\tilde{\vartheta})=\sum_{l=k-\hat{N}_{+}+1}^{k}  \|\boldsymbol{x}\left[l\right],\tilde{\boldsymbol{x}}_{\tilde{\vartheta}}\left[l\right]\|^2,
\end{equation}
\begin{equation}
\label{eq:min}
E\left[k\right]=\min_{\tilde{\vartheta}} {e(k,\tilde{\vartheta})},\quad
\vartheta=\argmin_{\tilde{\vartheta}} {e(k,\tilde{\vartheta})},
\end{equation}
where $\tilde{\boldsymbol{x}}_{\tilde{\vartheta}}$ is the state obtained from simulating \eqref{eq:sys} with excitation $\hat{\boldsymbol{u}}_{+}=g(\hat{\boldsymbol{u}},\tilde{\vartheta})$ and with initial state $x[k-\hat{N}_{+}]$. 
This value provides an estimate of how good the model would fit after a small update.

Provided the small update is sufficient (see Sec.~\ref{sec:lt-trigger}), the parameters $\vartheta$ are transmitted to the receiver and, subsequently, the receiver determines the new trajectory using the predefined function $g$ and the previous model trajectory $\hat{\boldsymbol{u}}$. For a cyclic process, useful parameters could be cycle length, phase shift, or amplitude, which can be estimated using standard signal processing methods in frequency or time domain to lower the computational costs in comparison to solving the optimization problem \eqref{eq:cost} explicitly (for an example see Sec.~\ref{sec:eval}). Learning of these parameters is always carried out as first step of the block \emph{model learning} in Fig.~\ref{fig:ETL}.

\subsection{Learning-type Trigger}
\label{sec:lt-trigger}
The small update described above requires much less communication than an update of the full trajectory $\hat{\boldsymbol{u}}$, but might not always lead to sufficiently precise predictions. 
We define a binary trigger $\gamma _{\text{full}}\left[k\right]\in \left\{0,1\right\}$ that indicates when a small update is not sufficient to achieve a satisfying performance improvement.
														
If the error $E\left[k\right]$ exceeds a threshold $\alpha \in \mathbb{R}_{0}^{+}$, then a full update is triggered
\begin{equation}
\label{eq:lt-trigger}
\gamma _{\text{full}}\left[k\right]=\begin{cases}
 0 & \text{if }\gamma _{\text{learn}}\left[k\right]=0\vee  E\left[k\right] \leq \alpha \\
1 & \text{if }\gamma _{\text{learn}}\left[k\right]=1\wedge E\left[k\right] >    \alpha 
\end{cases}
\end{equation}
and a new full trajectory $\hat{u}$ is identified and transmitted as detailed in Sec.~\ref{sec:traj}.
Otherwise ($\gamma _{\text{full}}\left[k\right]=0$), the sensor sends the parameters $\vartheta$ as small model update and both agents deform the previous excitation trajectory to obtain the new model $\hat{\boldsymbol{u}}=\hat{\boldsymbol{u}}_{+}$.
The triggers of both model updates and the state update exhibit a hierarchical dependency: A small model update is only executed if the (communication-wise cheaper) state updates occur too frequently. Likewise, a (communication-wise expensive) full model update is only carried out if a small model update is expected to be insufficient.

\subsection{Model Learning -- Full Model Update} \label{sec:traj}
\label{sec:full}

In case of a triggered full update, an appropriate new model $\hat{\boldsymbol{u}}$ cannot be obtained from the previous one by applying the deformation $g$. Therefore, the excitation trajectory for the previously measured cycle is estimated based on the available state measurements. We assume that the dynamics $f$ allow for estimating $u[k]$ from a finite number of state measurements (as is the case in the considered application, see Sec.~\ref{sec:eval}).

This trajectory could be transferred directly to the receiving agent via the network and used as a precise model by both prediction blocks for the following state estimations. However, sending all samples of the trajectory would require the network to transfer many values in a short time interval. Therefore, the sender compresses the trajectory at first with polynomial regression and the receiver performs reconstruction to obtain $\hat{\boldsymbol{u}}$. Possible alternatives to polynomial regression for compression are, e.g., wavelet transformation with thresholding of the coefficients \cite{wav}, and (sparse) Gaussian process regression \cite{gpr,sparseGpr}. 

The entire approach described in Sec. \ref{sec:ETL} is summarized in Algorithm~1.

\begin{subfigures}
\begin{figure}
\end{figure}
\begin{algorithm}
  \caption{ETL for  Cyclically Excited Systems (Sender)}
  \begin{algorithmic}
    \IfThenElse {$j < N$}
      {$j \gets j + 1$}
      {$j \gets 1$}  \begin{footnotesize} \Comment{Update index \eqref{eq:index}} \end{footnotesize} 
    \State $\hat{x} \gets f\left(\hat{x},\hat{u}_{j},0\right)$ \begin{footnotesize} \Comment{Predict measurement \eqref{eq:pred}} \end{footnotesize}
    \If {$\left\| x,\hat{x}\right\|<\delta$} \begin{footnotesize} \Comment{Prediction error small \eqref{eq:su-trigger}} \end{footnotesize}
    \State $t \gets t + 1$ \begin{footnotesize} \Comment{Increase inter-communication time} \end{footnotesize}
    \Else \begin{footnotesize} \Comment{Prediction error large \eqref{eq:su-trigger}}  \end{footnotesize}
    \State $\hat{x} \gets x$ \begin{footnotesize} \Comment{Update state with measurement \eqref{eq:pred}} \end{footnotesize}
    \State $\boldsymbol{\tau} \gets \left[\begin{array}{cc} \boldsymbol{\tau} & t \end{array} \right]$ \begin{footnotesize} \Comment{Store inter-communication time} \end{footnotesize}
    \State $t \gets 0$ \begin{footnotesize} \Comment{Reset inter-communication time} \end{footnotesize}
    \EndIf
    \IfThenElse {$\mathrm{KSTest}\left(\boldsymbol{\tau}\right)$}
      {$l \gets 0$}
      {$l \gets l + 1$} \begin{footnotesize} \Comment{KS-Test \eqref{eq:l-trigger}} \end{footnotesize}
    \State $\boldsymbol{x} \gets \left[\begin{array}{cc} \boldsymbol{x} & x \end{array} \right]$ \begin{footnotesize} \Comment{Store measurement} \end{footnotesize}
    \If {$l \geq t_{\mathrm{min}}$} \begin{footnotesize} \Comment{Minimum holding time reached \eqref{eq:l-trigger}} \end{footnotesize}
    \State $j \gets 0$ \begin{footnotesize} \Comment{Reset model trajectory index \eqref{eq:index}} \end{footnotesize}
    \State $\boldsymbol{\tau} \gets \emptyset$ \begin{footnotesize} \Comment{Empty inter-communication time buffer} \end{footnotesize}
    \State $\vartheta \gets \mathrm{estimateParam}\left(\boldsymbol{x}\right)$ \begin{footnotesize} \Comment{Small update parameter \eqref{eq:cost}, \eqref{eq:min}} \end{footnotesize}
    \State $\hat{\boldsymbol{u}} \gets g\left(\hat{\boldsymbol{u}}, \vartheta\right)$ \begin{footnotesize} \Comment{Small model update (Sec.~\ref{sec:small})} \end{footnotesize}
    \If {$\mathrm{trajError}\left(\hat{\boldsymbol{u}},\boldsymbol{x}\right)>\alpha$} \begin{footnotesize} \Comment{Small update poor \eqref{eq:min}, \eqref{eq:lt-trigger}} \end{footnotesize}
    \State $\hat{\boldsymbol{u}} \gets \mathrm{identifyU}\left(\boldsymbol{x}\right)$ \begin{footnotesize} \Comment{Full model update (Sec.~\ref{sec:full})} \end{footnotesize}
    \EndIf
    \EndIf
  \end{algorithmic}
\end{algorithm}
\end{subfigures}

\section{Application: Pitch Angle of Human Foot}  \label{sec:eval}

To demonstrate the effectiveness of the proposed algorithms, we consider a network of wearable sensor units comprising at least an inertial sensing chip, a wireless communication module, and a microcontroller, which is attached to a human body during gait. Such measurement systems are used for real-time biofeedback and control of robotic systems and neuroprostheses \cite{neuroControl}. The rate at which the network can communicate reliably in real time is limited by the number of sensors. If the communication load between each sensor and the receiver can be reduced, higher base sampling rates or a larger number of sensors can be used. This challenge was recently addressed using heuristic approaches \cite{stop}. We now apply the proposed ETL methods to this problem.

We consider the specific example of real-time measurement of the foot pitch angle $x\left[k\right]\in \mathbb{R}$ in a feedback-controlled neuroprostheses \cite{neuroControl} (cf. Fig.~\ref{fig:ETSE}). The angle dynamics is modeled by cyclic increments $u\left[k\right]\in \mathbb{R}$ and zero-mean additive Gaussian noise $\varepsilon \left[k\right]\sim \mathcal{N}\left(0,\sigma^2 \right)$ with standard deviation $\sigma \in \mathbb{R}_{0}^{+}$, i.e., the state update is
\begin{equation}
\label{eq:appSys}
x\left[k\right]=x\left[k-1\right]+u\left[k\right]+\varepsilon \left[k\right].
\end{equation}
Equation \eqref{eq:appSys} corresponds to \eqref{eq:sys} and is used for state predictions. For the trigger \eqref{eq:su-trigger}, the absolute difference $d\left[k\right]=\left|x\left[k\right]-\hat{x}\left[k\right]\right|$ is used as a metric.

Validation is based on measurements of approximately 30~minutes of variable human gait including normal walking with frequent speed and ground inclination changes as well as style changes to different kinds of simulated pathological gait (straightened knee, drop foot, and dragged leg). While walking style changes lead to clearly different angle trajectories, small gait velocity changes are well described by time warping of the angle trajectory, i.e., by cycle length changes of the periodic excitation $u\left[k\right]$. The small model update is, therefore, designed as follows: Determine the new cycle length $\vartheta_{1}=\hat{N}$ and time shift $\vartheta_{2}$ of $\hat{\boldsymbol{u}}$ that best describe the measurements of the most recent cycle. Both parameters are estimated in the frequency domain (using the auto- or crosscovariance of the measured states of a small number of previous cycles) and further refined in the time domain with local optimization\footnote[3]{Code an detailed description on \url{http://www.control.tu-berlin.de/EventTriggeredLearning}}. 


For a full model update, \eqref{eq:appSys} with $\varepsilon[k]$ being zero-mean implies that an unbiased estimate of $u[k]$ can be obtained by taking the difference of $x[k]$ and $x[k-1]$. We thus calculate $\Delta x[l] := x[l]-x[l-1]~\forall~ l\in[k-\hat{N}+1,k]$ and estimate the full trajectory as $\hat{\boldsymbol{u}} = \left[\begin{array}{ccc} \Delta x[k-\hat{N}+1] & \cdots & \Delta x\left[k\right] \end{array} \right]$, where $k$ is the index of the sampling instant where learning is triggered.


\subsection{Parametrization} \label{sec:param}

The measurement sampling rate is set to 50 Hz. The accuracy of IMU-based foot pitch angle measurements is in the range of 3--4$^{\circ}$ \cite{accuracy}. Therefore, the maximum admissible angle error $\delta $ in \eqref{eq:su-trigger} is set to 2$^{\circ}$. The significance level for the KS-test \eqref{eq:l-trigger} is set to $\eta =5\;\% $. The accompanying minimum holding time is $t_{\min }=0.35\;\mathrm{s}\cdot50\;\mathrm{Hz}$, which is about a third of a stride duration. The initial MC simulation to obtain $F(\tau )$ is carried out with 1000 trials and a standard deviation $\sigma =0.9^{\circ} $ of the noise. This value is the observed root-mean-square-error (RMSE) during normal walking.
The RMSE-threshold of the learning-type trigger \eqref{eq:lt-trigger} is $\alpha =5^{\circ} $.
Polynomial regression is chosen for compression of $\hat{\boldsymbol{u}}$. The degree is fixed to 18, which leads to a maximum of 20 communicated parameters including the constant offset and the estimated cycle length $\hat{N}$. In contrast, the full trajectory of a stride typically contains 40--80 samples if the rate is 50 Hz. More parameters would increase the risk of overfitting.

The MC simulation yields an expected inter-communication time $\hat{\mathbb{E}}\left[\tau \right]\approx 8$. This implies that a state update is expected to happen every 8$^{\mathrm{th}}$ sample on average if the model is correct. Furthermore, because of $\eta =5\;\% $, every 20$^{\mathrm{th}}$ state update is expected to trigger an undesired model update, which leads to transmission of either two or 20 model parameters. In total, the parametrization should result in not more than $\frac{V_{\mathrm{ETL}}}{V_{\text{full}}}\leq \frac{n+\eta w }{n\cdot\mathbb{\hat{E}}\left[\tau \right]} = \frac{1 + 0.05 \cdot 20}{8} = 25 \; \%$ of transmitted 
values in comparison to full communication if the model is correct (i.e., whenever gait style and velocity do not change).

Finally, we initialize the model $\hat{\boldsymbol{u}}$ of the excitation trajectory with a vector containing zeros. The first estimated model is transmitted when the first full model update is triggered.

\subsection{Results and Discussion} \label{sec:res}

\begin{figure}[t]
      \centering
      \includegraphics[width=\columnwidth]{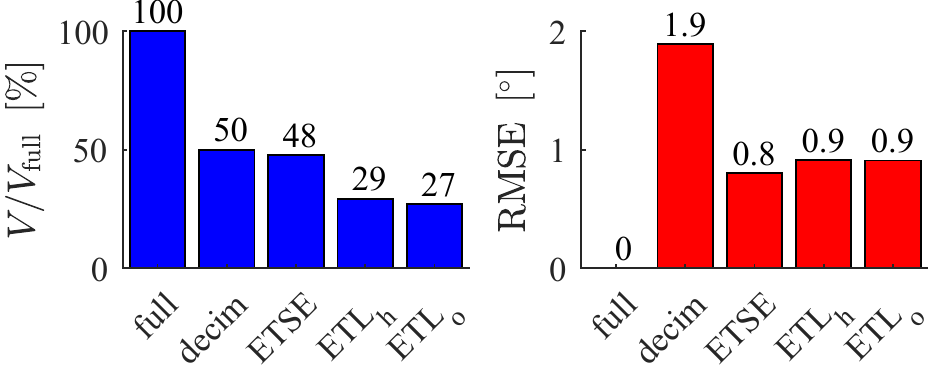}
      \caption{Number of transferred values $V$ with respect to full communication $V_{\text{full}}$ (left) and resulting RMSE (right). We compare full communication (full), decimation with factor 2 (decim), event-triggered state estimation (ETSE), and two parametrizations of event-triggered learning (ETL$_\text{h}$ / ETL$_\text{o}$) for about 0.5 h of highly variable gait. The results show that ETL reduces the communication significantly without changing the accurancy of the transferred signal in comparison to ETSE.}
      \label{fig:bars}
   \end{figure}

The proposed ETL method is compared to alternative strategies by determining the number of transmitted values and the RMSE between the measured angle and the output of the receiver. Results are visualized in Fig.~\ref{fig:bars}. The trivial strategy of sending only every second sample (decim.) yields an RMSE of almost 2$^{\circ}$ at 50 \% communication rate. The same reduction of communication is achieved by pure ETSE (i.e., the scheme in Fig.~\ref{fig:ETL}, in black without ETL); however, at an RMSE of less than 1$^{\circ}$. ETL achieves a reduction of traffic to even less than 30 \%, while the RMSE is still below 1$^{\circ}$. Finally, optimizing the heuristically chosen parameters (ETL$_{\text{h}}$) with nested cross-validation (ETL$_{\text{o}}$) leads to small further improvements (27 \%).

\begin{figure}[thpb]
      \centering
      \includegraphics[width=\columnwidth]{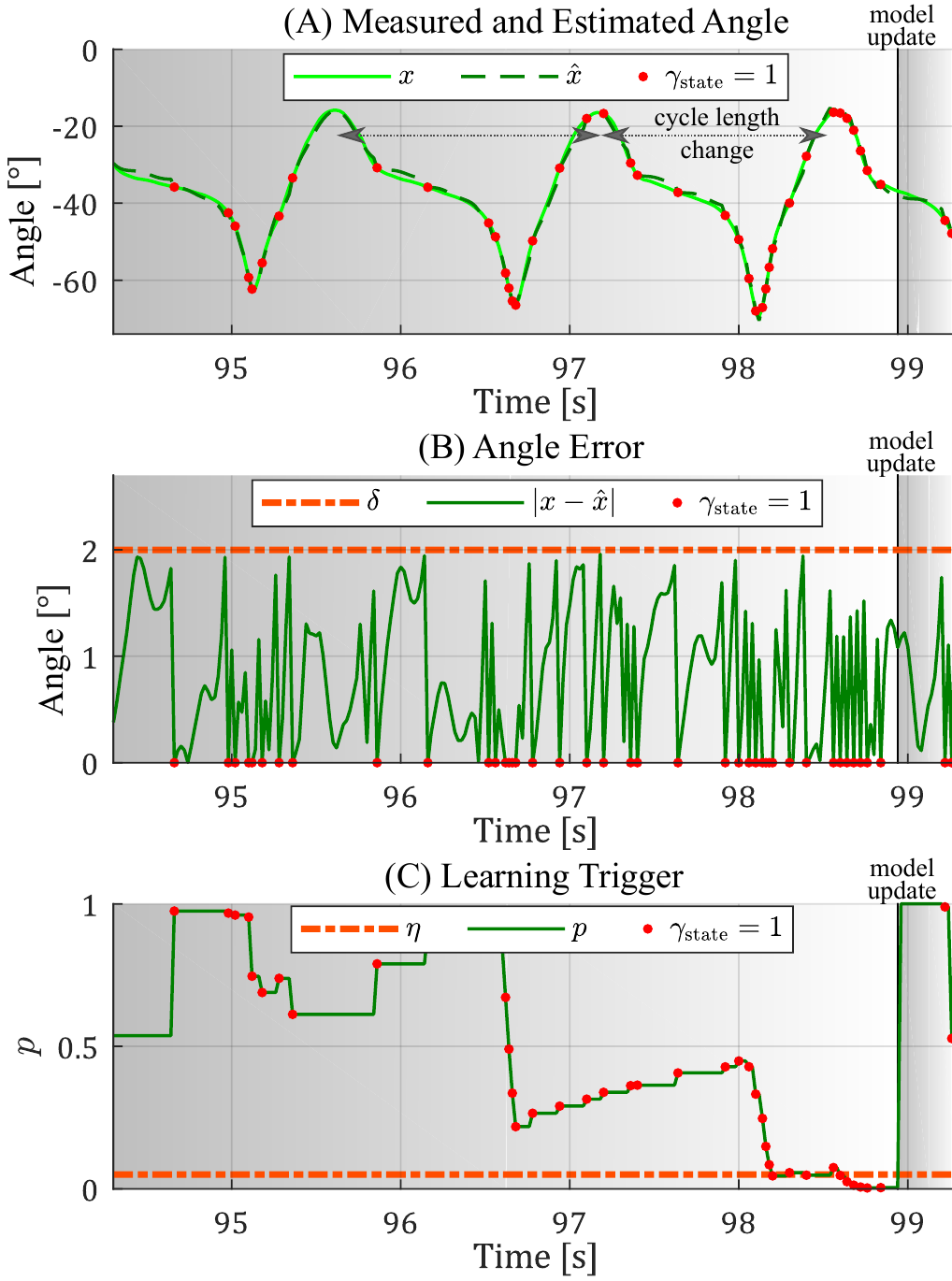}
      \caption{Small model update ($\gamma _{\text{learn}}[k]=1\wedge \gamma _{\text{full}}[k]=0$) triggered at 98.9~s due to a change of cycle length (gait velocity). (A) measured and estimated state; (B) error $\left|x\left[k\right]-\hat{x}\left[k\right]\right|$ and state-update trigger threshold $\delta$; (C) probability $p\left[k\right]$ of the KS-test \eqref{eq:l-trigger} and significance level $\eta $. Model learning is triggered because too many state updates ($\gamma _{\text{state}}[k]=1)$ occur and $p\left[k\right]$ falls below $\eta $ for the minimum holding time $t_{\text{min}}$.}
      \label{fig:small}
   \end{figure}

\begin{figure}[thpb]
     \centering
     \includegraphics[width=\columnwidth]{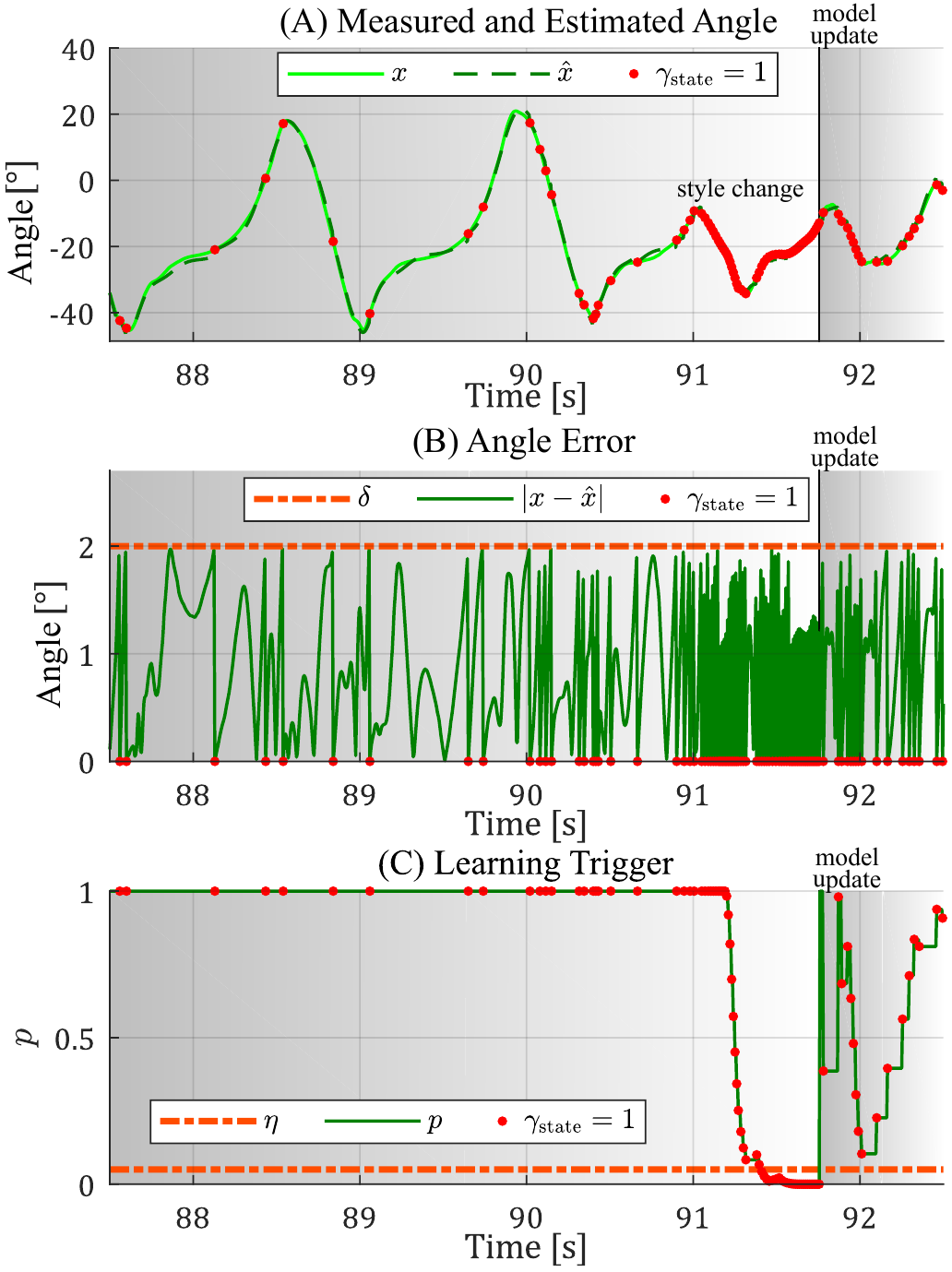} 
     \caption{Full model update ($\gamma _{\text{learn}}[k]=1\wedge \gamma _{\text{full}}[k]=1$) triggered at 91.8~s due to a change of trajectory shape (walking style). Plotted signals and quantities are the same as in Fig.~\ref{fig:small}.}
   \label{fig:full}
  \end{figure}

Fig.~\ref{fig:small} shows how a small model update is triggered due to a change of the cycle length of the measured process, while Fig.~\ref{fig:full} demonstrates how a change of the shape of the trajectory causes a full model update. Both figures also provide evidence that the absolute estimation error is bounded by $\delta =2^{\circ} $. This represents a remarkable improvement with respect to the results in \cite{stop}.

\section{Conclusions} \label{sec:conclusions}




A set of methods for event-triggered learning for cyclically excited systems is proposed to minimize communication in sensor networks. The approach automatically recognizes cyclic patterns in data -- even when they change repeatedly -- and reduces communication load whenever the current data can be accurately predicted from previous cycles. In contrast to previous approaches, the current methods exploit explicitly the periodicity of the dynamics and account for time-variant behavior. Additional major advantages of the methods are that they assure an upper bound on the error of the received signal and that unnecessary model updates occur only with a low probability, which is controlled by a user-defined significance level. 

The proposed methods are 
shown to yield significant resource savings in a wireless body sensor network for orientation tracking of a human foot. 
The experimental results show that a large reduction of communication load (by 70 \%) and a small bounded estimation error (by 2$^{\circ}$) can be achieved at the same time. This means that up to three times as many sensors could be used without jeopardizing latency of the real-time communication. Moreover, the examined sampling rate of 50 Hz is relatively low; up to 1 kHz is common in IMU networks (200 Hz in wireless ones), which gives even more potential for resource savings with the proposed methods.

Future research will focus on quaternion-based estimation of full body segment orientations and a learning trigger with less assumptions on inter-communication times.

\bibliographystyle{IEEEtran}
\bibliography{refs}

\end{document}